\documentclass{iau} 
\usepackage{graphicx}

\title[Exploring molecular complexity in the Galactic Center with ALMA] 
{Exploring molecular complexity in the Galactic Center with ALMA}

\author[Arnaud Belloche]   
{Arnaud Belloche$^1$}

\affiliation{$^1$Max-Planck-Institut f\"ur Radioastronomie, Auf dem H\"ugel 69, 53121 Bonn, Germany
\\ email: {\tt belloche@mpifr-bonn.mpg.de}} 

\pubyear{2017}
\volume{332}  
\jname{Astrochemistry VII -- Through the Cosmos from Galaxies to Planets}
\editors{M. Cunningham, T. Millar \& and Y. Aikawa, eds.}
\begin{document}

\maketitle

\begin{abstract}
The search for complex organic molecules (COMs) in the ISM has revealed 
chemical species of ever greater complexity. This search relies heavily on the 
progress made in the laboratory to characterize the rotational spectra of 
these molecules. Observationally, the advent of ALMA with its high angular 
resolution and sensitivity has allowed to reduce the spectral confusion and 
detect low-abundance molecules that could not be probed before. We present 
results of the EMoCA survey conducted with ALMA toward the star-forming region 
Sgr~B2(N). 
This spectral line survey aims at deciphering the molecular content of 
Sgr~B2(N) in order to test the predictions of astrochemical models and gain 
insight into the chemical processes at work in the ISM. We report on the 
tentative detection of N-methylformamide, on deuterated COMs, and on the 
detection of a branched alkyl molecule. Prospects for probing molecular 
complexity in the ISM even further are discussed at the end.

\end{abstract}

\firstsection 
\section{Complex organic molecules in the interstellar medium}
\label{s:intro}

The in-situ exploration of comet 67P/Churyumov-Gerasimenko by the Rosetta
mission has led to the detection of more than twenty organic molecules in 
the coma of the comet and in material excavated from the surface by the impact 
of Rosetta's lander Philae (\cite{Goesmann15,Altwegg16,Altwegg17}). Most of 
these molecules, such as methyl cyanide, CH$_3$CN, formamide, NH$_2$CHO, or 
acetone, CH$_3$C(O)CH$_3$, were already known to exist in the interstellar 
medium (ISM) but some, like methyl isocyanate, CH$_3$NCO, reported by 
\cite{Goesmann15}\footnote{The identification of methyl isocyanate in the mass
spectrum of the cometary material analyzed by \cite{Goesmann15} has been 
recently disputed (\cite{Ligterink17,Altwegg17}).}, were not. Shortly after, 
as soon as spectroscopic 
predictions of the rotational spectrum of the molecule became available, 
methyl isocyanate was detected in the ISM (\cite{Halfen15,Cernicharo16}).
Another molecule that was not known to exist in the ISM, and has still not 
been found, is the amino acid glycine, NH$_2$CH$_2$COOH. \cite{Altwegg16} 
reported its identification in the coma of 
comet 67P, confirming an earlier claim by \cite{Elsila09} who analyzed samples 
returned from comet 81P/Wild~2 to Earth by the Stardust mission.
In addition to these cometary detections, more than 80 different types of 
amino acids were identified in meteorites on Earth, with an isotopic 
composition and a racemic distribution suggesting that their presence is not
due to contamination by Earth material and that they have instead a true 
extraterrestrial origin (see, e.g., \cite{Botta02}).
The degree of molecular complexity found in these small solar-system bodies, 
in particular in meteorites, raises the following questions: is this chemical 
complexity a widespread outcome of interstellar chemistry in our Galaxy, 
and possibly even in other galaxies? What degree of complexity is interstellar 
chemistry capable of achieving, in particular in star (and planet) forming 
regions? 

\begin{figure}
\begin{center}
\includegraphics[width=0.8\hsize]{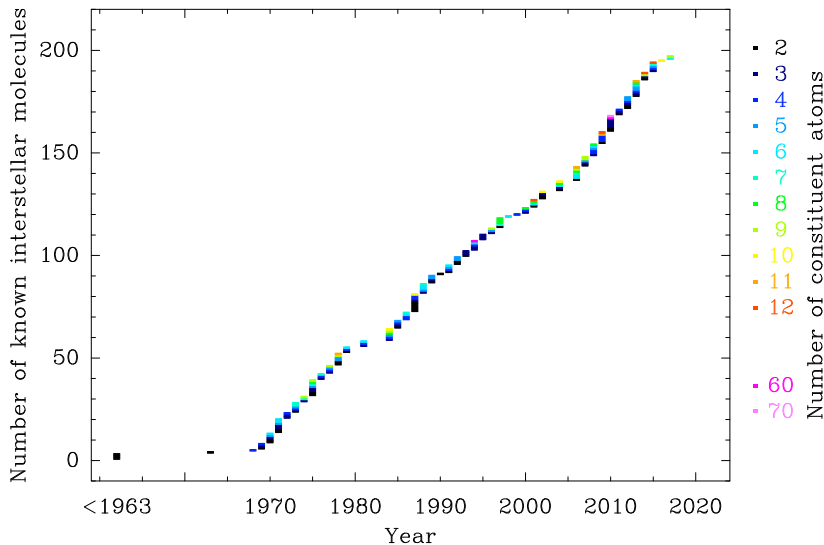}
\end{center}
\caption{Number of known interstellar molecules as a function of time as of
July 2017. The three molecules identified before 1963 are 
indicated on the left. The number of atoms each molecule contains is 
color-coded following the color scheme shown on the right. This diagram uses
the information provided by H.~S.~P. M\"uller in the Cologne Database for 
Molecular Spectroscopy (CDMS, http://www.astro.uni-koeln.de/cdms/molecules).}
\label{f:spacemol}
\end{figure}

As of today, 197 molecules have been identified in the interstellar medium or 
in circumstellar envelopes of evolved stars. This number does not include 
isotopologs, that is molecules containing less abundant isotopes such as
D, $^{13}$C, $^{18}$O, $^{34}$S, or $^{15}$N. The first identifications of 
interstellar molecules, CH, CH$^+$, and CN, go back nearly 
eight decades ago and were made in absorption in the UV domain 
(\cite{McKellar40,Adams41,Douglas41}). However, the search for molecules in 
the ISM started to become efficient more than two decades later, with the 
advent of radio astronomy. Figure~\ref{f:spacemol} shows the number of known 
interstellar molecules as a function of time, with the number of their 
constituent atoms color-coded. Since the first detection of a molecule in the 
radio domain, the hydroxyl radical OH by \cite{Weinreb63}, about seven 
molecules have been newly identified in the ISM every two years on average.
While the rate of detection has been to first order constant over the past 
five decades, it is worth mentioning that the detection rate over the past 
decade ($\sim$5 new molecules per year) exceeds the rate achieved during the 
first decade when radio astronomy started to flourish ($\sim$4 new molecules 
per year between 1968 and 1977). This shows the current vitality of this field 
of research and suggests that the inventory of interstellar molecules should 
continue to grow significantly in the near future.

\begin{figure}
\begin{center}
\includegraphics[width=0.9\hsize]{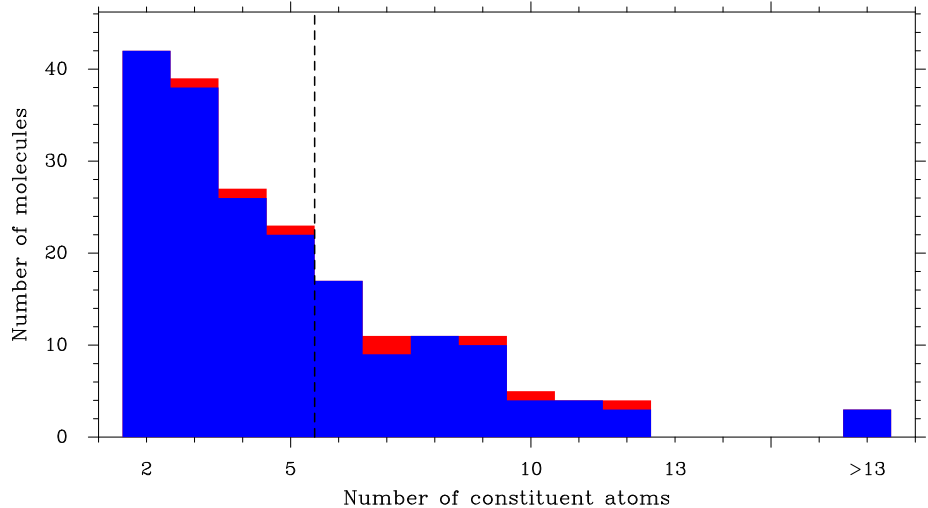}
\end{center}
\caption{Number of known interstellar molecules as a function of the number of 
their constituent atoms as of July 2017. Detections 
published in 2015--2017 are emphasized in red. The dashed line indicates the 
threshold above which a molecule is defined as complex.}
\label{f:histo}
\end{figure}

Following the review by \cite{Herbst09}, it has become common in the 
astrochemical community to define a molecule as complex when it contains six 
atoms or more. With this definition, about one third of the known interstellar
molecules are complex
(Fig.~\ref{f:histo}). All these complex molecules contain at least one carbon 
atom and they are thus called complex organic molecules (COMs). The presence 
of such complex interstellar molecules raises the question of their origin. 
What are the chemical processes that lead to chemical complexity in the ISM? 
What are the respective roles of gas-phase and grain-surface/mantles 
processes? What are the relevant reactions? What are their reaction rates and
branching ratios? To answer all these questions, a close interplay between 
laboratory experiments, theoretical calculations, numerical simulations, and 
observations is necessary (see, e.g., \cite{Herbst09,Garrod13b,Oberg16}). Here, 
we focus on the latter with the aim of confronting results of observations to 
predictions of astrochemical models.

\section{Sgr~B2 and the EMoCA survey}
\label{s:emoca}

\begin{figure}
\begin{center}
\includegraphics[width=0.7\hsize]{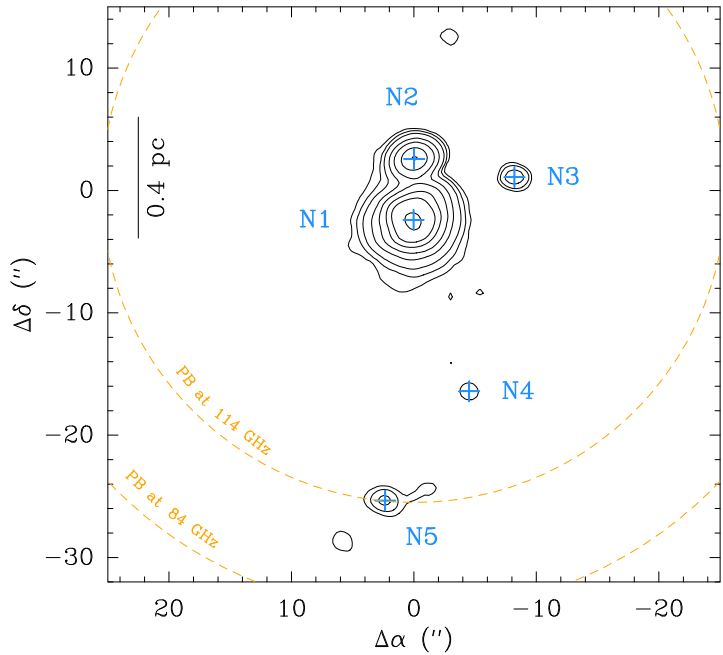}
\end{center}
\caption{Contour map of Sgr~B2(N) representing the number of channels between 
84.1~GHz and 114.4~GHz with emission detected above the $7\sigma$ level in the 
continuum-subtracted spectra of the EMoCA survey. The contour levels are 500, 
1000, 2000, 5000, 10000, and then increase by steps of 10000. The dashed 
circles represent the primary beam of the ALMA 12~m antennas at 84~GHz and 
114~GHz. The equatorial offsets are given with respect to the phase center 
located at ($\alpha, \delta$)$_{\rm J2000}$ =
($17^{\rm h}47^{\rm m}19.87^{\rm s}, -28^\circ22'16''$).
Sources identified as hot cores by \cite{Bonfand17} are indicated with blue 
crosses and labels. Figure adapted from \cite{Bonfand17}.}
\label{f:linecountmap}
\end{figure}

One of the most prominent star-forming regions in our Galaxy is the giant 
molecular cloud Sagittarius (Sgr) B2, with a mass of $\sim$10$^7$~M$_\odot$ in
a diameter of $\sim$40~pc (\cite{Lis90}). Sgr~B2 is located about 100~pc from 
the Galactic Center in projection, at a distance of 8.3~kpc from the Sun 
(\cite{Reid14}). It contains two main sites of on-going formation of high-mass 
stars, Sgr~B2(N) and Sgr~B2(M), that both host clusters of ultra-compact 
H\,\textsc{\lowercase{II}} regions (see, e.g., \cite{Gaume95,Schmiedeke16}).
Sgr~B2(N), in particular, contains several hot molecular cores 
(Fig.~\ref{f:linecountmap}). The main ones, Sgr~B2(N1) and Sgr~B2(N2), have 
been known for a long time (e.g., 
\cite{Snyder94,Belloche08,Qin11,Belloche16}), but several new, fainter ones 
were recently detected with the Atacama Large Millimeter/submillimeter Array 
(ALMA) (\cite{Bonfand17,SanchezMonge17}). The H$_2$ column densities of these 
hot cores are so high (10$^{24}$--10$^{25}$~cm$^{-2}$ over few arcsec) that the 
detection of molecules with low abundances in Sgr~B2 is easier than toward 
other sources. This has been a key advantage for the detection of new COMs in 
the past few decades. Indeed, many COMs were first detected toward Sgr~B2. 

Following up a spectral line survey of Sgr~B2(N) and Sgr~B2(M) carried out 
with the IRAM~30~m single-dish telescope a decade ago, which led to the 
detection of several new COMs (\cite{Belloche08,Belloche09,Belloche13}), we
performed the EMoCA project with ALMA in Cycles 0 and 1. EMoCA stands for 
Exploring Molecular Complexity with ALMA. This interferometric project is a 
spectral line 
survey of Sgr~B2(N) that covers the 3~mm atmospheric window between 
84.1~GHz and 114.4~GHz, at an angular resolution of $\sim$1.6'' and
a spectral resolution of 488~kHz, that is a velocity resolution between 1.3
and 1.7~km~s$^{-1}$. We achieved a sensitivity of $\sim$3~mJy~beam$^{-1}$ 
($1\sigma$), that is $\sim$0.15~K in brightness temperature scale.
EMoCA aims at detecting new COMs in the ISM and, thereby, testing the 
predictions delivered by numerical simulations of interstellar chemistry.

\begin{figure}
\begin{center}
\includegraphics[width=1.0\hsize]{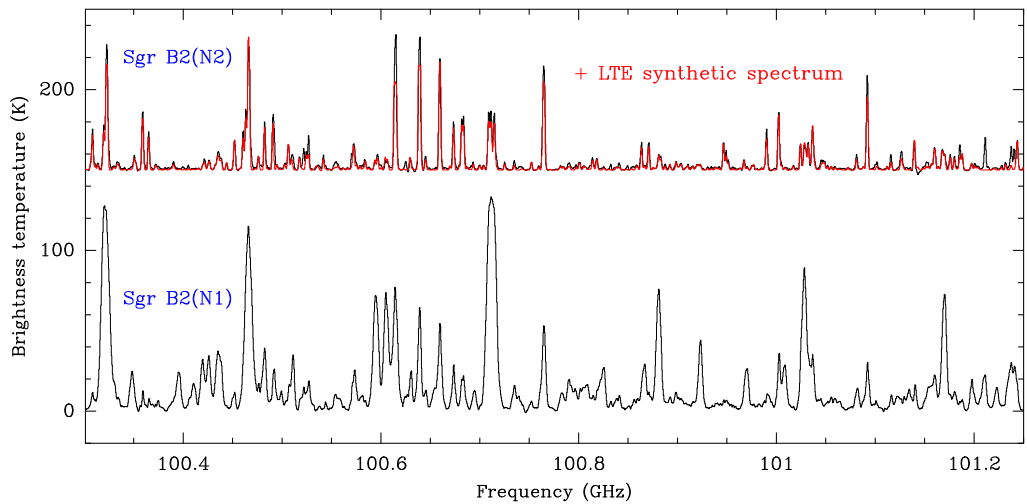}
\end{center}
\caption{Short portion of the EMoCA spectrum observed toward the hot molecular 
cores Sgr~B2(N1) and Sgr B2(N2) with ALMA at an angular resolution of 
$\sim$1.6$''$. The red spectrum shows the LTE model that contains the 
contribution of all the molecules identified so far toward Sgr B2(N2).}
\label{f:specn1n2}
\end{figure}

In addition to delivering maps of thousands of spectral lines, the EMoCA 
survey represents an improvement by more than one order of magnitude both in
angular resolution and sensitivity compared to our previous single-dish 
survey. In particular, its angular resolution of 1.6'' (13000 au in 
projection) is sufficient to separate the two main hot cores, Sgr~B2(N1) and 
Sgr~B2(N2), that are separated by $\sim$5'' in the north-south direction. A 
nice surprise of this survey is that the emission lines of the northern hot 
core, Sgr~B2(N2), have widths (FWHM) of 5~km~s$^{-1}$ only, which reduces 
considerably the level of line confusion compared to previous single-dish data.
As shown in Fig.~\ref{f:specn1n2}, the lines detected toward the main hot core,
Sgr~B2(N1), are broader ($\sim 7$~km~s$^{-1}$) and some of them have stronger 
wings. With the aim to look for weak lines of low-abundance species, we have
focused our analysis on Sgr~B2(N2) so far. We model its emission spectrum
using Weeds (\cite{Maret11}) under the assumption of local thermodynamic 
equilibrium (LTE), which is reasonable given the high densities probed with 
ALMA. Our spectroscopic database, which is needed to model the spectra, 
contains all entries available in the JPL and CDMS databases 
(\cite{Pickett98,Mueller01,Endres16}), as well as additional predictions sent 
to us by spectroscopists. Our current model of the emission spectrum of 
Sgr~B2(N2) is shown in red in Fig.~\ref{f:specn1n2}.
In the next three sections, we report on some of the results obtained 
with the EMoCA survey based on this model.

\section{Tentative detection of N-methylformamide}
\label{s:ch3nhcho}

N-methylformamide, CH$_3$NHCHO, is a structural isomer of the known 
interstellar molecule acetamide, CH$_3$C(O)NH$_2$, that was detected in 
Sgr~B2(N) by \cite{Hollis06} (see also \cite{Halfen11}). Both molecules contain
a peptide bond which is the characteristic chemical bond linking amino acids 
in proteins. N-methylformamide can also be seen as a partially hydrogenated
form of methyl isocyanate, CH$_3$NCO, which was recently detected in the ISM
as mentioned in Sect.~\ref{s:intro}. Given that both acetamide and methyl 
isocyanate were known interstellar molecules, N-methylformamide appeared as
a good candidate for an interstellar detection. New measurements of the 
rotational spectrum of the \textit{trans} conformer of this molecule were 
performed in Lille (France) and Kharkiv (Ukraine) over the frequency range
45--630~GHz, and accurate spectroscopic predictions were then obtained 
with the RAM36 program (\cite{Ilyushin10}) for its ground state and its first 
two torsionally excited states (\cite{Belloche17}).

\begin{figure}
\begin{center}
\includegraphics[width=1.0\hsize]{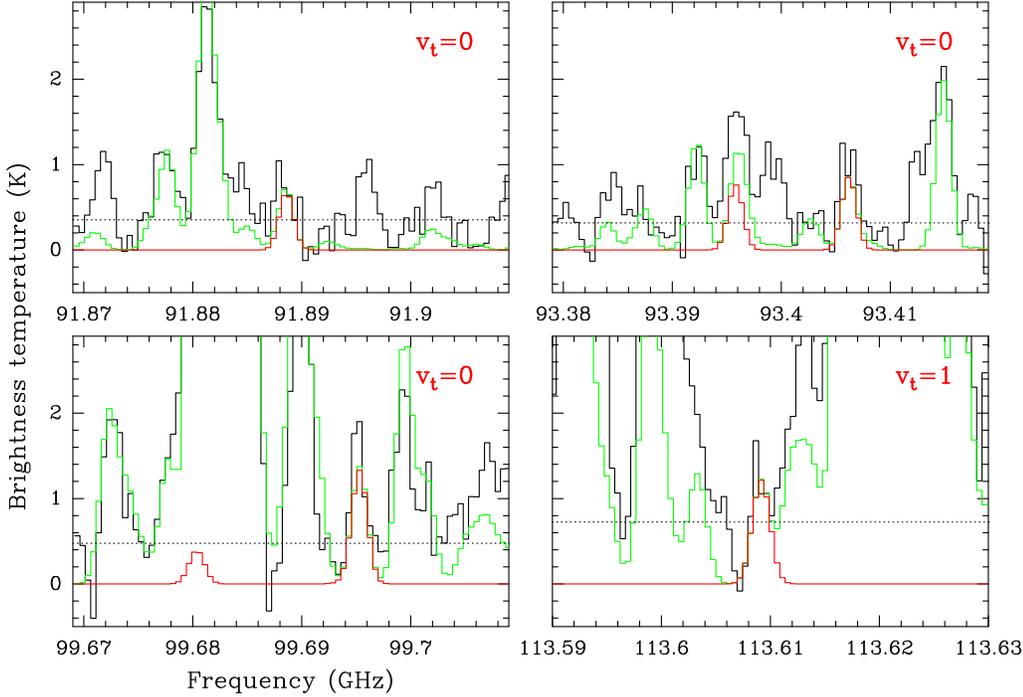}
\end{center}
\caption{Spectra of Sgr~B2(N2) showing four of the five transitions of 
N-methylformamide that do not suffer much from contamination by emission lines
of other species in the EMoCA survey (\cite{Belloche17}). In each panel, the 
observed ALMA spectrum is shown in black, the synthetic spectrum of 
N-methylformamide in red, 
and the model containing the contributions of all molecules identified so 
far in green. The dotted line indicates the $3\sigma$ detection level. The 
line in the bottom right panel belongs to the first torsionally excited state 
of N-methylformamide, while the other ones are ground-state lines.}
\label{f:spec_ch3nhcho}
\end{figure}

With these spectroscopic predictions, we searched for emission of 
N-methylformamide in the EMoCA spectrum of Sgr~B2(N2). Five of the lines that
we assigned to CH$_3$NHCHO suffer little from contamination by emission
lines of other species. Four of these five lines are shown in 
Fig.~\ref{f:spec_ch3nhcho}. Given that our full model of Sgr~B2(N2) includes 
the contribution of all species identified so far (red spectrum in 
Fig.~\ref{f:specn1n2} and green spectrum in Fig.~\ref{f:spec_ch3nhcho}), we 
selected seven additional lines of N-methylformamide that could be 
``decontaminated'' with the full model and we used them together with the five 
well detected lines to build a population diagram for this molecule. A fit to 
this diagram yields a temperature of $\sim$150~K, which is typical of the 
rotational 
temperatures derived for other COMs in Sgr~B2(N2). Since only five lines are
detected without contamination, the detection of CH$_3$NHCHO is considered
as tentative only (\cite{Belloche17}). However, the fact that our model takes
into account the emission of all species identified so far and that the 
derived rotational temperature of N-methylformamide is typical of COM emission 
in Sgr~B2(N2), we are confident that the identification of this molecule in 
this source is correct.

We used the EMoCA survey to compare the abundance of N-methylformamide to the
ones of related species. CH$_3$NHCHO is, within a factor of two,  
nearly as abundant as CH$_3$NCO and CH$_3$C(O)NH$_2$, and more than one order
of magnitude less abundant than formamide, NH$_2$CHO, and isocyanic acid, HNCO.
We compared these abundance ratios to predictions obtained with the chemical 
kinetics model MAGICKAL (\cite{Garrod13a}), the network of which was expanded
to include N-methylformamide as well as related species. The first conclusion
of this comparison is that an efficient formation of HNCO via NH+CO on grains
with an activation energy below 1500~K is necessary to achieve the observed 
abundance of CH$_3$NCO. The second conclusion is that the 
production of CH$_3$NHCHO on grain surfaces is plausible via two formation 
routes: the successive addition of H to CH$_3$NCO and radical-radical reactions 
(CH$_3$+HNCHO or HCO+HNCH$_3$) (\cite{Belloche17}). The good match between
the predicted and measured abundance ratios is an additional support to our
identification of N-methylformamide in the EMoCA spectrum of Sgr~B2(N2). 
However, the chemical model produces an overabundance of acetamide, which
should be investigated in more details in the future.

\section{Deuterated complex organic molecules}

The cosmic abundance of deuterium with respect to hydrogen is low 
($\sim$$1.5\times10^{-5}$, \cite{Linsky03}). Still, molecules in star-forming
regions are often found to be enriched in deuterium (e.g., \cite{Parise06}).
This mainly results from the exothermicity of the reaction of H$_3^+$  with
HD, the reservoir of deuterium in molecular clouds, that produces H$_2$D$^+$, 
and the depletion of CO, the main destroyer of H$_3^+$, onto the surface of 
dust grains that occurs in the dense and cold regions of molecular clouds.
Given these properties, the degree of deuterium fractionation of molecules in 
star-forming regions is a powerful tool to trace the history of their cold 
prestellar phase (see, e.g., \cite{Caselli12,Ceccarelli14}). 
Although the level of deuteration is known to be low in the Galactic Center 
region (e.g., \cite{Gerin92}), we took advantage of the high
sensitivity of the EMoCA survey to search for deuterated COMs in Sgr~B2(N2).

\begin{figure}
\begin{center}
\includegraphics[width=0.58\hsize]{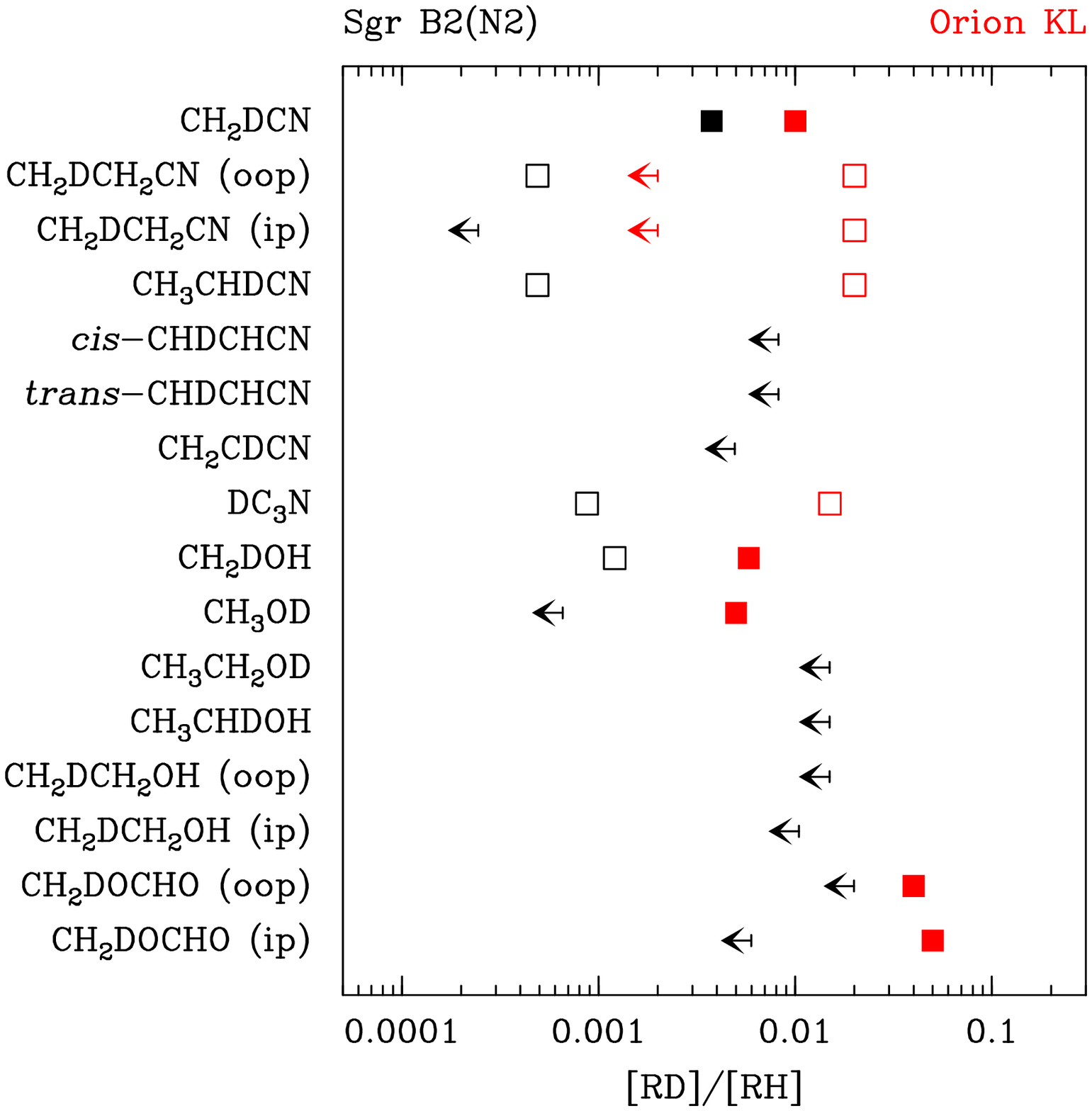}\hspace*{0.02\hsize}\includegraphics[width=0.40\hsize]{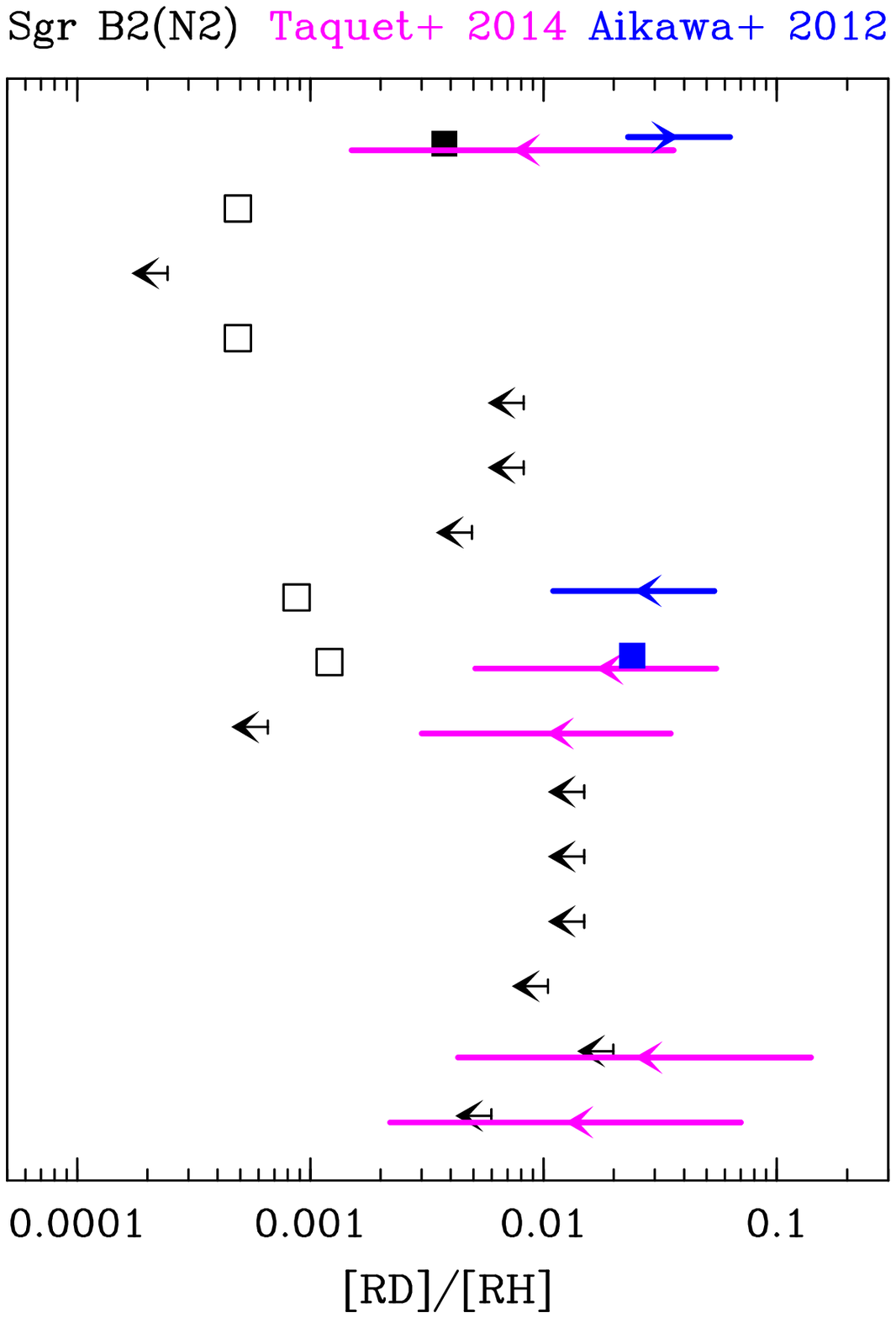}
\end{center}
\caption{\textit{Left panel:} Level of deuterium fractionation measured for a
sample of (complex) organic molecules toward Sgr B2(N2) (\cite{Belloche16}). 
Filled and empty squares indicate robust and tentative detections, 
respectively. Upper limits are indicated with arrows. The symbols in red show 
the level of deuteration reported for Orion~KL by 
\cite{Gerin92,Margules09,Daly13,Esplugues13,Neill13}, and \cite{Coudert13}.
\textit{Right panel}: Same measurements toward Sgr~B2(N2) compared to 
predictions of chemical models of collapsing protostellar envelopes. The purple
bars and arrows show the evolution of the level of deuterium fractionation as
a function of time during the main accretion phase in the model of 
\cite{Taquet14} while the blue ones and the blue square show the model 
predictions of \cite{Aikawa12}.}
\label{f:dfrac}
\end{figure}

We firmly detected CH$_2$DCN and could also report tentative detections of 
CH$_2$DCH$_2$CN, CH$_3$CHDCN, DC$_3$N, and CH$_2$DOH (\cite{Belloche16}). The
level of deuteration measured for these species, as well as upper limits
derived for other deuterated COMs, are summarized in the left panel of 
Fig.~\ref{f:dfrac}, along with measurements or upper limits reported in the
literature for Orion~KL. Overall, the degree of deuteration of COMs in 
Sgr~B2(N2) is on the order of $0.1\%$, a factor three to ten lower than in 
Orion~KL. The right panel of Fig.~\ref{f:dfrac} compares the Sgr~B2(N2) 
results to predictions of chemical models of (low-mass) protostellar envelopes 
(\cite{Aikawa12,Taquet14}). Apart for a possible agreement for CH$_2$DCN in
one model, the models produce overabundances of deuterated COMs by at least a 
factor of four.

The low level of deuteration of COMs in Sgr~B2(N2) has at least two possible 
origins. First, it may reflect a lower elemental abundance of deuterium in the 
Galactic Center region compared to the local ISM, owing to stellar processing 
of previous generations of stars, as has been claimed, albeit with large 
uncertainties, by previous studies that were based on small molecules 
(\cite{Jacq99,Lubowich00,Polehampton02}). A second possibility could be the 
thermal history of star formation in Sgr~B2. Both the gas kinetic temperature 
(30--50~K, \cite{Ott14}) and the dust temperatures (20--28~K, \cite{Guzman15}) 
are higher in Sgr~B2 than in nearby star forming regions. In the model of 
\cite{Taquet14}, the level of deuteration of methanol is reduced by a factor 
three at the end of the prestellar phase when the initial dust temperature is 
increased by 3~K only, from 17~K to 20~K. The deuteration of COMs is thus very 
sensitive to the temperature during the prestellar phase, and the higher 
temperatures prevailing in the Galactic Center region could have contributed 
to the low level of COM deuteration found in Sgr~B2(N2). In addition, it is 
worth mentioning that the factor five difference found in Sgr~B2(N2) between 
the deuteration level of methyl cyanide, CH$_3$CN, and ethyl cyanide, 
C$_2$H$_5$CN, may point to different formation pathways, in the gas phase for 
the former and in the ice mantles for the latter. 
Dedicated numerical simulations of deuteration designed for Sgr B2 will be 
needed in the future to investigate its thermal history and the chemical 
pathways of its deuterium chemistry in detail.

\section{Branched molecules in the interstellar medium}
\label{s:iprcn}

\begin{figure}
\begin{center}
\includegraphics[width=1.0\hsize]{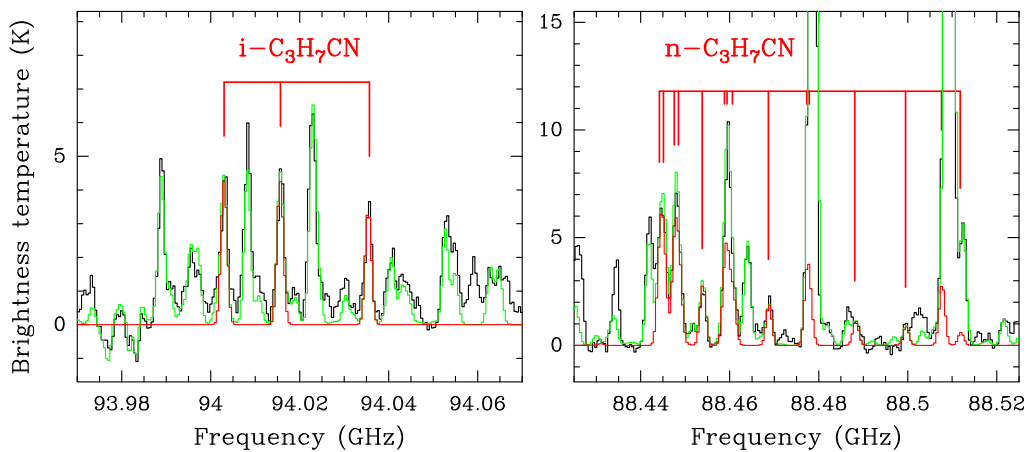}
\end{center}
\caption{\textit{Left panel:} Example of transitions of \textit{iso}-propyl 
cyanide detected toward Sgr~B2(N2) with ALMA (\cite{Belloche14}). The observed 
spectrum is shown 
in black, the synthetic spectrum of \textit{iso}-propyl cyanide in red, and the 
model containing the contributions of all molecules identified so far in green.
\textit{Right panel}: Example of transitions of \textit{normal}-propyl 
cyanide detected toward Sgr~B2(N2) with ALMA. The red spectrum is the 
synthetic spectrum of \textit{normal}-propyl cyanide.}
\label{f:spec_prcn}
\end{figure}

One of the first results of the EMoCA survey was the interstellar detection of 
a branched alkyl molecule (\cite{Belloche14}). The rotational spectrum of this 
molecule, \textit{iso}-propyl cyanide, \textit{i}-C$_3$H$_7$CN, was studied in 
the laboratory by \cite{Mueller11}. It is a structural isomer of the 
straight-chain form \textit{normal}-propyl cyanide, \textit{n-}C$_3$H$_7$CN,
that we detected earlier in our single-dish survey of Sgr~B2(N) 
(\cite{Belloche09}). Both isomers are clearly detected in the EMoCA spectrum, 
with about 50 transitions for the branched form and 120 transitions for the
chain-like form (Fig.~\ref{f:spec_prcn}). The population diagram of 
\textit{n-}C$_3$H$_7$CN yields a well constrained rotational temperature
of $\sim$150~K, typical of the COM emission in Sgr~B2(N2). From the LTE
modelling, we derived a surprisingly high abundance ratio 
[\textit{i-}C$_3$H$_7$CN]/[\textit{n-}C$_3$H$_7$CN] of 0.4, with the branched
form nearly as abundant as the chain-like one.

We compared the measured abundance ratio of the two isomers of propyl cyanide 
again to predictions of the chemical kinetics model MAGICKAL
(\cite{Garrod13a}). The reaction network of this model was expanded to include, 
for the first time, the formation of branched molecules 
(\cite{Garrod17,Belloche14}). The measured abundance ratio of the two isomers 
of propyl cyanide is well reproduced 
by the model. The model indicates that both molecules can be formed on grain 
surfaces, but via different formation routes. The dominant routes for 
\textit{n-}C$_3$H$_7$CN are the addition reactions CH$_3$\.CH$_2$+\.CH$_2$CN 
and \.CN+CH$_2$=CHCH$_3$ while the formation of \textit{i-}C$_3$H$_7$CN is
dominated by the radical-radical reaction \.CH$_3$+CH$_3$\.CHCN. In the model, 
the key parameter influencing the \textit{i}/\textit{n} ratio is the barrier 
of the \.CN addition to C$_2$H$_2$ and C$_2$H$_4$ (\cite{Garrod17}).

As mentioned in Sect.~\ref{s:intro}, amino acids of extraterrestrial origin 
were detected in meteorites on Earth. Both branched and chain-like isomers of 
these amino acids were identified, and the branched isomers were even found to 
dominate over the chain-like ones (e.g., \cite{Cronin83}). The detection of a
branched alkyl molecule in the ISM thus establishes a further link between
the chemical composition of meteorites and interstellar chemistry.

\section{Outlook}

The EMoCA project is just one example illustrating the new window opened by 
ALMA in terms of angular resolution and sensitivity to probe the complexity of
interstellar chemistry and confront the derived chemical composition of 
astronomical sources to the predictions of astrochemical models. While Sgr~B2 
is a place of choice to detect molecules of low abundance thanks to its high 
column densities, spectral line surveys of other sources with ALMA, such as 
the PILS survey toward a 
nearby low-mass protostar (The ALMA Protostellar Interferometric Line Survey, 
\cite{Jorgensen16}), are exploring the outcome of interstellar chemistry
in other environments and will also contribute to improving our 
understanding of the chemical processes at work in the ISM.

Our study of deuterated COMs in Sgr~B2(N2) has shown that a sensitive, 
unbiased spectral line survey at high angular resolution is an 
exquisite tool to measure the level of deuterium fractionation of molecules 
with high accuracy. This is made possible by the conditions prevailing in this 
hot core: the densities are so high that collisions dominate over radiative 
processes and the LTE approximation is valid. The spectra of COMs, which can 
have several dozens of lines detected, can thus be fitted with great accuracy. 
In this respect, Sgr~B2 is an excellent laboratory to test the predictions of 
chemical models quantitatively. Numerical simulations of deuterium chemistry
designed for Sgr~B2(N) should help us in the future set valuable constraints
on the D/H elemental abundance in the Galactic Center region.

After the detection of acetamide, CH$_3$C(O)NH$_2$, and the tentative detection
of N-methylformamide, CH$_3$NHCHO (Sect.~\ref{s:ch3nhcho}), it is tempting to
search for the next structural isomer of these two molecules, acetimidic acid,
CH$_3$C(OH)NH, which is the third most stable isomer in the C$_2$H$_5$NO
family (\cite{Lattelais10}). However, the rotational spectrum of this 
molecule has not been characterized in the laboratory yet, and its dipole 
moment is a factor of two smaller than the ones of acetamide and 
N-methylformamide (\cite{Lattelais09}). Therefore, its detection in the ISM 
will be challenging.

\begin{figure}
\begin{center}
\includegraphics[width=0.27\hsize]{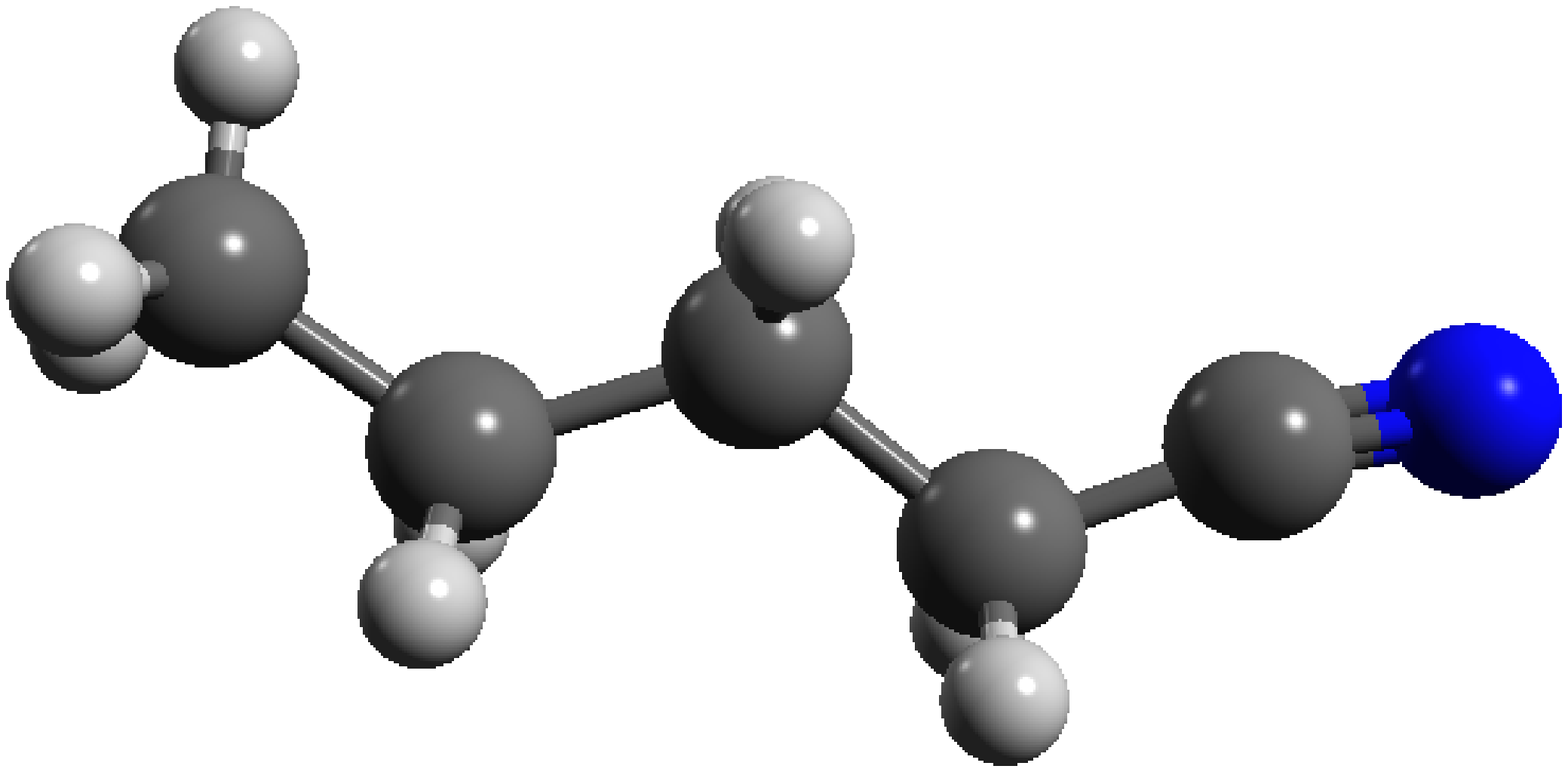}\hspace*{0.03\hsize}\includegraphics[width=0.17\hsize]{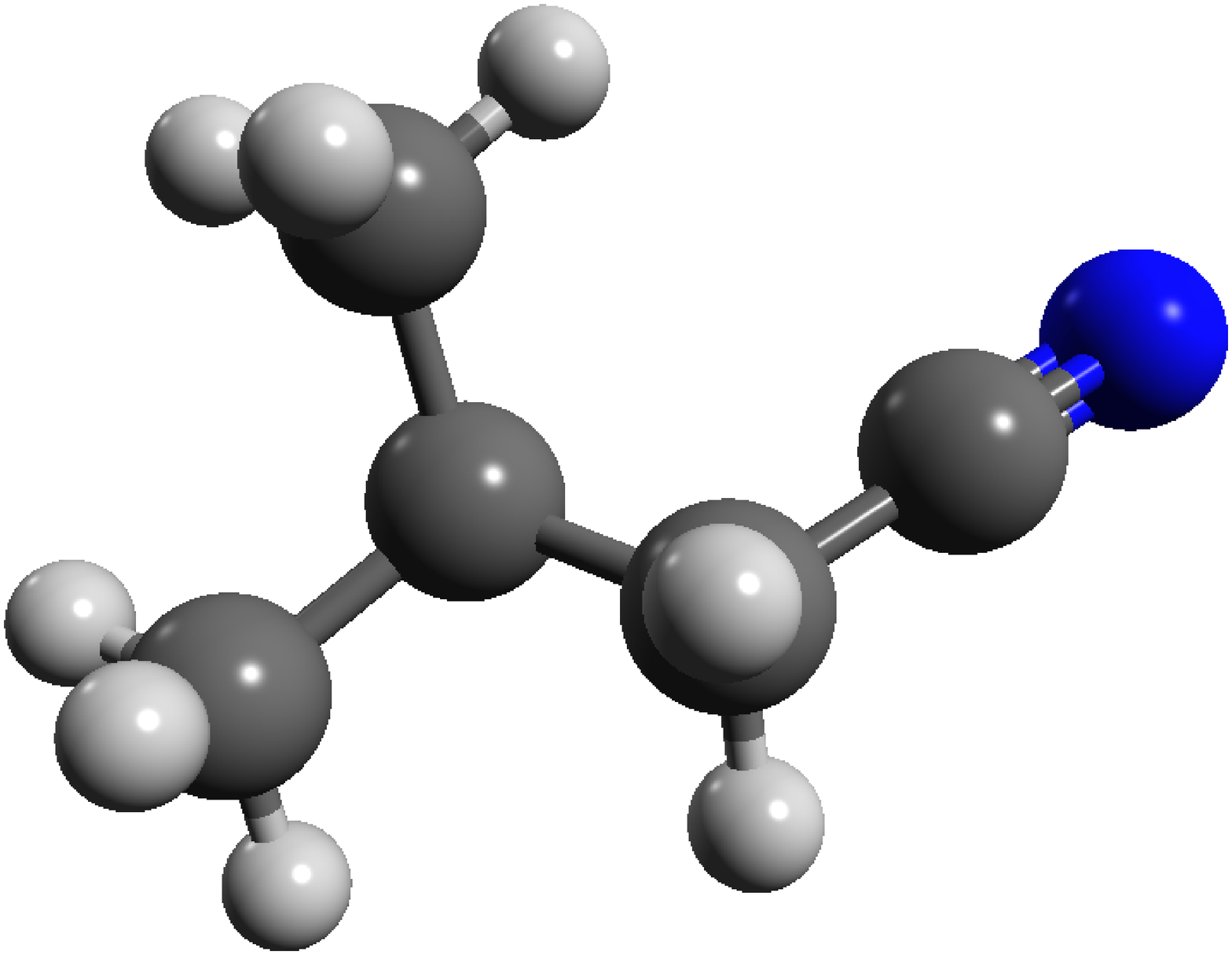}\hspace*{0.03\hsize}\includegraphics[width=0.22\hsize]{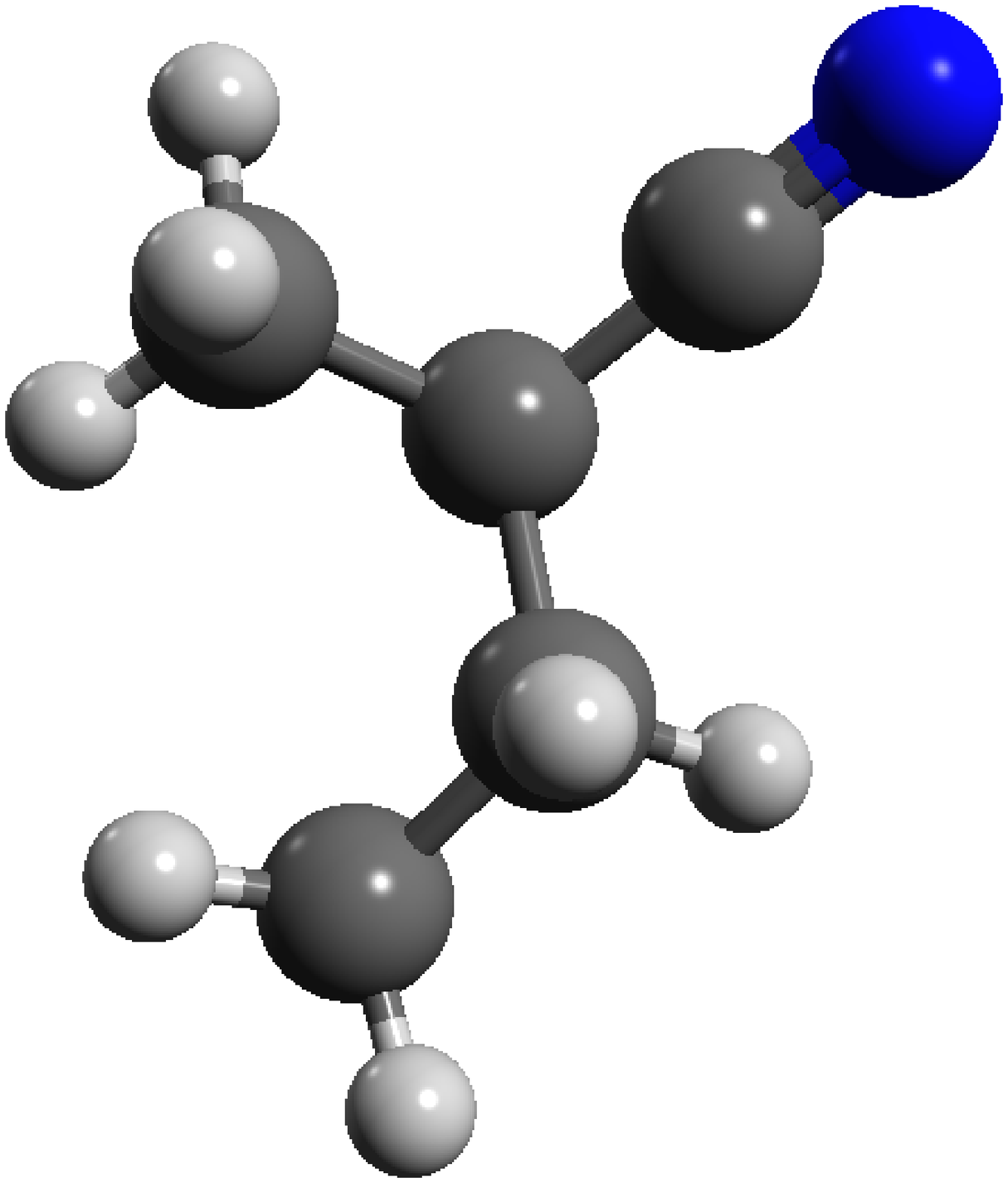}\hspace*{-0.01\hsize}\includegraphics[width=0.23\hsize]{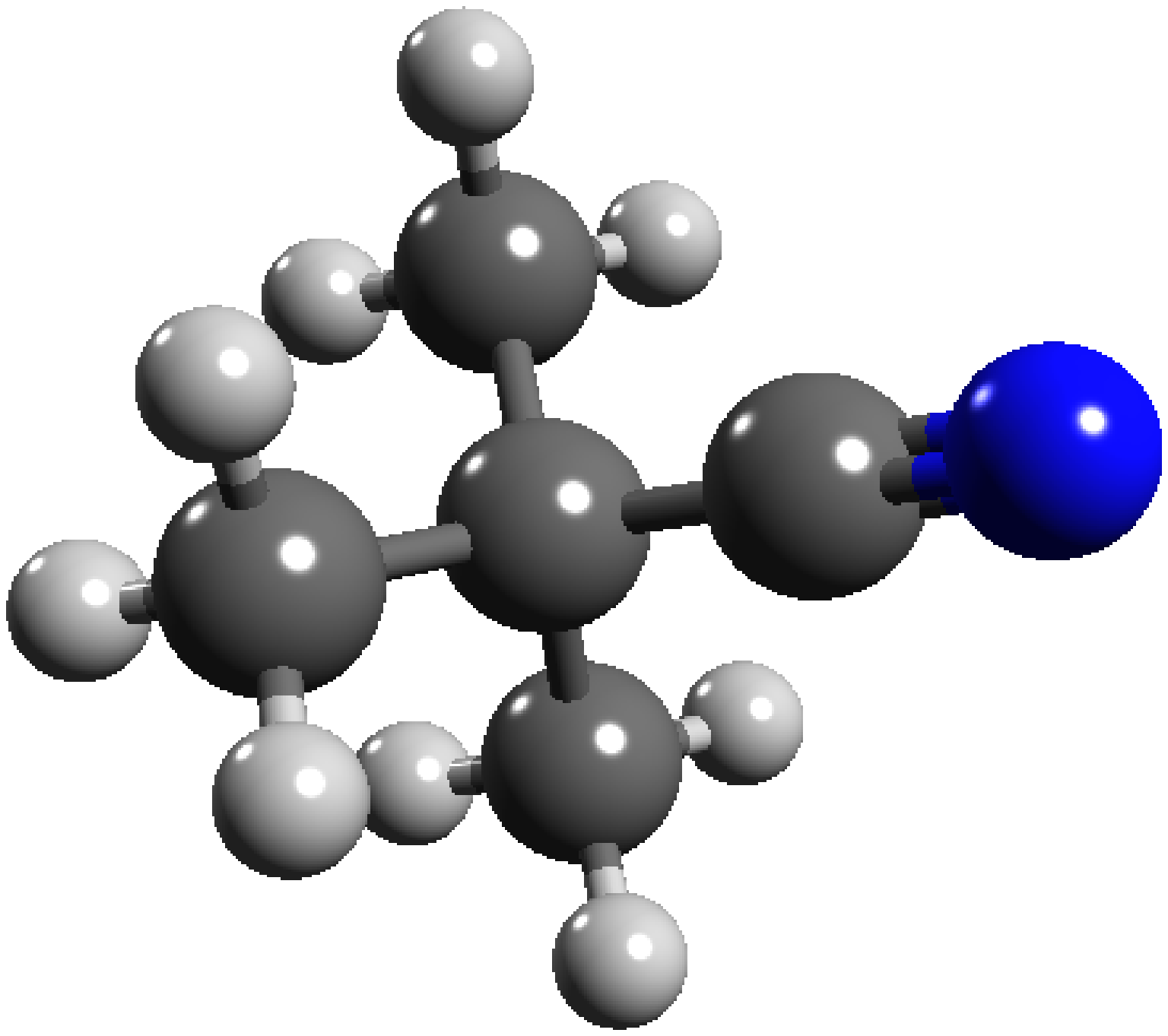}
\end{center}
\caption{Ball-and-stick representations of the chain-like and branched forms 
of butyl cyanide. From left to right: \textit{n}-C$_4$H$_9$CN, 
\textit{i}-C$_4$H$_9$CN, \textit{s}-C$_4$H$_9$CN, and \textit{t}-C$_4$H$_9$CN. 
Figure adapted from \cite{Garrod17}, image credit: E. Willis.}
\label{f:bucn}
\end{figure}

The detection of a branched alkyl molecule (Sect.~\ref{s:iprcn}) has opened a 
new domain in the structures available to the chemistry of star forming 
regions. The next step in complexity in the alkyl cyanide family is butyl
cyanide, C$_4$H$_9$CN. This molecule has one straight-chain isomer and three 
branched ones (Fig.~\ref{f:bucn}). The straight-chain form, 
\textit{normal}-butyl cyanide, was searched for in the ISM but it has not 
been detected so far, neither in our previous IRAM 30~m survey of Sgr~B2(N) 
nor in the EMoCA survey (\cite{Ordu12,Garrod17}). The chemical model MAGICKAL 
predicts that one of the branched forms, \textit{s}-C$_4$H$_9$CN, should in
fact dominate over the straight-chain form in Sgr~B2(N2) and that we may have
been only short of a detection with the EMoCA survey (\cite{Garrod17}). This
motivated us to start a follow-up survey of Sgr~B2(N) at 3~mm with ALMA in its 
Cycle 4. This new project will represent an improvement by a factor of three 
both in angular resolution and sensitivity with respect to the EMoCA survey. 
The characterization of the rotational spectrum of \textit{s}-C$_4$H$_9$CN was 
performed in Cologne (\cite{Mueller17}) and the Cycle 4 observations have been 
completed recently. \textit{s}-C$_4$H$_9$CN will be one of the first molecules
we will search for in this new, sensitive spectral line survey that may 
reveal the presence of other interesting COMs as well.

As mentioned in Sect.~\ref{s:emoca}, the EMoCA survey revealed that the 
hot core Sgr~B2(N2) has linewidths as narrow as 5~km~s$^{-1}$ at arcsecond
scale. This helped reduce the spectral confusion compared to previous 
single-dish surveys. EMoCA has indeed not reached the confusion limit on 
this source yet, while the spectrum of Sgr~B2(N1) is almost confusion-limited. 
There is thus still room for improvement on Sgr~B2(N2), and also on the other 
fainter hot cores detected in Sgr~B2(N) (\cite{Bonfand17,SanchezMonge17}), 
which is one of the goals of our new Cycle 4 project. However, it is 
likely that the confusion limit will almost be reached in this next survey and
the following question will arise: how will we beat the confusion limit in 
order to explore even further the chemical complexity of the ISM in the
future? At least two possibilities can be investigated. As spectral confusion 
increases with frequency in Sgr B2(N) (\cite{Belloche13}), the first option is 
to go to lower frequencies, as has already been explored at low angular 
resolution with the PRIMOS spectral survey of Sgr B2(N) performed with the 
Green Bank Telescope (e.g., \cite{Neill12}). Higher angular resolution will 
however be necessary to probe the hot cores in an optimal way: this will be 
possible with ALMA once it is equipped with receivers operating below 80~GHz 
(bands 1 and 2 or 2+3, \cite{DiFrancesco13,Fuller16}), and on a longer term 
with the Next Generation Very Large Array (ngVLA, \cite{Hughes15}). The second 
direction will be to perform very deep integrations with ALMA on sources that 
have narrower linewidths, such as the Class 0 protostar IRAS~16293--2422 which 
is the target of the PILS survey (\cite{Jorgensen16}).

\vspace*{2ex}
\noindent\textit{Acknowledgements:} I would like to thank Holger M\"uller, Rob 
Garrod, Karl Menten, M\'elisse Bonfand, Vivien Thiel, and Eric Willis for 
their precious contributions to the EMoCA project. I am also thankful to the 
whole spectroscopic community for tirelessly providing spectroscopic
predictions of molecules of ever greater complexity to the astronomical
community. This paper makes use of the following ALMA data:\\
ADS/JAO.ALMA\#2011.0.00017.S, ADS/JAO.ALMA\#2012.1.00012.S. 
ALMA is a partnership of ESO (representing its member states), NSF (USA), and 
NINS (Japan), together with NRC (Canada), NSC and ASIAA (Taiwan), and KASI 
(Republic of Korea), in cooperation with the Republic of Chile. The Joint ALMA 
Observatory is operated by ESO, AUI/NRAO, and NAOJ. 
This work has been in part supported by the Deutsche Forschungsgemeinschaft 
(DFG) through the collaborative research grant SFB 956 ``Conditions and Impact 
of Star Formation'', project area B3.

\end{document}